\documentclass[prl, reprint,showpacs,  superscriptaddress]{revtex4-1}

\usepackage{graphicx}
\usepackage{dcolumn}
\usepackage{graphicx}
\usepackage{epsfig}
\usepackage{epsf}
\usepackage{amssymb,amsmath,amsthm,bm}
\usepackage{empheq,cases}
\usepackage{mathtools}
\usepackage{multirow}
\usepackage[colorlinks=true,linkcolor=blue!50!black,citecolor=red,pdfauthor={ },pdftitle={ },pdfsubject={ },pdfkeywords={ }]{hyperref}
\usepackage{tikz}
\usepackage{pgfplots}
\usetikzlibrary{calc}
\usepackage{lipsum}

\usepackage{times}

\theoremstyle{definition}

\definecolor{c1}{RGB}{254,232,200}
\definecolor{c2}{RGB}{253,187,132}
\definecolor{c3}{RGB}{227,74,51}

\usepackage{multirow}
\usepackage{dcolumn}
\newcolumntype{d}[1]{D{.}{.}{#1}}

\begin{document}

\title{Subsystem-based Approach to Scalable Quantum Optimal Control}

\author{Jun Li}
\email{lij3@sustech.edu.cn}
\affiliation{Institute for Quantum Science and Engineering and Department of Physics, Southern University of Science and Technology, Shenzhen 518055, China}
%\affiliation{Center for Quantum Computing, Peng Cheng Laboratory, Shenzhen 518055, China}

\begin{abstract}
The development of  quantum control methods is an essential task for emerging quantum technologies.
In general, the  process of optimizing  quantum controls   scales  very unfavorably  in system size due to the exponential growth of the  Hilbert space dimension.   Here, I present a scalable subsystem-based method for   quantum optimal control on a large quantum system. The basic idea is to cast  the   original problem as a robust  control  problem on subsystems with requirement of   robustness in respect  of inter-subsystem couplings, thus enabling a  drastic  reduction of problem size. The method   is in particular suitable for  target quantum operations that are local, e.g., elementary quantum gates.  As an illustrative example, I employ it  to tackle the demanding task of  pulse searching  on  a 12-spin coupled system,   achieving   substantial   reduction of    memory and time costs. This work has significant implications for  coherent quantum engineering on near-term    intermediate-scale quantum devices.
\end{abstract}

\maketitle

Precise and complete control   of multiple coupled quantum systems  plays a significant role in     the development of modern quantum science and engineering \cite{BCR10,Letokhov07,DAlessandro08,DP10,Acin18}. In particular, achieving control with high fidelity lies at the heart of enabling scaled quantum information processing. Quantum optimal control (QOC)  is a subject aimed at      finding a control law of optimal performance in steering  the dynamics of a quantum system to   some desired goal. Because of its ability to produce high-accuracy quantum operations,    QOC  has  become a standard tool   for various experimental platforms, ranging from nuclear magnetic resonance (NMR) \cite{Khaneja05,Kuprov11}, nitrogen-vacancy centers in diamond \cite{Jorg14,PhysRevLett.117.170501}, ultracold atoms \cite{PhysRevA.88.021601}, to superconducting qubits \cite{,PhysRevLett.112.240504,PhysRevLett.114.200502,PhysRevLett.120.150401} and trapped ions \cite{PhysRevLett.112.190502}.   
In the past decades,   a great effort has been put in developing fast and practical QOC methods   \cite{PhysRevLett.113.010502,PhysRevLett.112.240503,PhysRevLett.122.020601,PhysRevLett.120.150401,BCS17}. 
It turns out that, apart from certain simple cases where analytical solutions are available \cite{PhysRevA.63.032308,PhysRevA.65.032301,PhysRevLett.104.083001},  QOC in general has to exploit  numerical optimization techniques. 
A variety of   optimization algorithms have been developed and demonstrated to be powerful in solving QOC problems, to name a few, gradient ascent pulse engineering (GRAPE) \cite{Khaneja05},   differential evolution \cite{PhysRevLett.114.200502}, machine learning \cite{PhysRevX.8.031086}, etc. They have their respective advantages in respect of  algorithmic simplicity, convergence speed, or numerical stability. However,  there is a critical difficulty of scalability that limits their success   to only small-sized quantum devices. The process of control optimization often relies on heavy computational simulations of the dynamics of the system  under control. The fact that the dimensionality of the tensor product Hilbert space grows exponentially with  the system size can render an   optimal control search algorithm computationally intractable. Currently, there have been put forward only a few strategies attempting  to address this issue,   including  the use of tensor-network-based techniques \cite{PhysRevLett.106.190501,ma2018optimal} and the hybrid quantum-classical approach \cite{PhysRevLett.118.150503,Dawei17}.

 %The criteria for a good numerical QOC method are  generality,   flexibility and feasibility. More specifically, the primary concerns are computational issues such as convergence speed, algorithmic simplicity, numerical stability, etc.     

In this work, I propose a  simple  subsystem-based   approach to large quantum system optimal control. In this approach, the entire system is  divided   into a group of subsystems, and the entire system Hamiltonian is accordingly decomposed as intra-subsystem Hamiltonian plus inter-subsystem Hamiltonian. As constraints of   subsystem-based   optimization,  the control pulse should implement    all corresponding target operations of the  subsystems with high fidelity, and meanwhile needs to be robust  to      the couplings between the  subsystems.  In general, a pulse that is designed on the subsystems without consideration for the inter-subsystem couplings does not necessarily implement the desired unitary on the whole system. The essential point here is that, if the pulse has robustness, then  its global control fidelity   should deviate  only slightly with respect to its fidelities on the subsystems.  It is noted that the idea of   subsystem-based QOC has been previously considered in Ref. \cite{PhysRevA.78.012328}, but  the approach adopted therein   lacks   a solid theoretical basis  and thus is quite empirical.  Here,   I develop a methodology   in  transforming  a large-sized QOC problem into a collective robust control problem on the subsystems. Furthermore, I   show that the robust control problem can be effectively solved by a modified GRAPE algorithm under the Van Loan differential equation framework that was recently developed in Ref. \cite{HPZC19}. 
%While  a better-connected system of qubits may facilitate running quantum algorithms and quantum circuits, it is harder to synthesize controls \cite{Monroe17}.  For a general circuit, reducing a fully connected system to the more sparse star-shaped or linear nearest-neighbor connectivity requires an increase in the number of gates of $O(n)$, where $n$ is the number of qubits. 

%The next natural step is to develop robust, high-fidelity and scalable gates for larger scale quantum computation. The implementation of many-body interactions has been considered in various physical systems for quantum information including ion traps, atoms in optical lattices, and cold polar molecules.

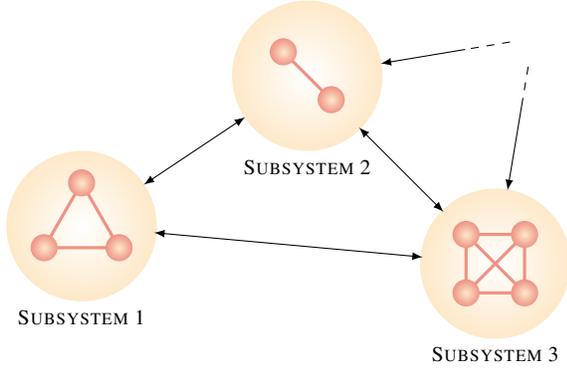
\begin{figure}[t]
\begin{tikzpicture}
\def\l{0.575}
\def\r{0.18}
	\draw [draw opacity=0,outer color=c1,inner color=white] (0,0) circle (1);
	\draw [c3!60,very thick] (0,\l) to (-0.866*\l,-0.5*\l) to (0.866*\l,-0.5*\l) to (0,\l);
	\draw [draw opacity=0,outer color=c3!60,inner color=c1] 
 (0,\l) circle (\r);
\draw [draw opacity=0,outer color=c3!60,inner color=c1] (-0.866*\l,-0.5*\l) circle (\r);
\draw [draw opacity=0,outer color=c3!60,inner color=c1](0.866*\l,-0.5*\l) circle (\r);
%	\node at (0,\l) {\sffamily\tiny Q1};
%	\node at (-0.866*\l,-0.5*\l) {\sffamily\tiny Q2};	
%	\node at (0.866*\l,-0.5*\l) {\sffamily\tiny Q3};

\def\l{0.45}	
	\draw [draw opacity=0,outer color=c1,inner color=white] (3,2) circle (1);
	\draw [c3!60,very thick] (3-0.707*\l,2+0.707*\l) to (3+0.707*\l,2-0.707*\l);
\draw [draw opacity=0,outer color=c3!60,inner color=c1](3-0.707*\l,2+0.707*\l) circle (\r);	
\draw [draw opacity=0,outer color=c3!60,inner color=c1](3+0.707*\l,2-0.707*\l) circle (\r);	
%	\node at (3-0.707*\l,2+0.707*\l) {\sffamily\tiny Q4};
%	\node at (3+0.707*\l,2-0.707*\l) {\sffamily\tiny Q5};	
	
\def\l{0.55}		
	\draw [draw opacity=0,outer color=c1,inner color=white]  (5.5,-0.5) circle (1);	
	\draw [c3!60,very thick] (5.5-0.707*\l,-0.5+0.707*\l) to (5.5-0.707*\l,-0.5-0.707*\l) to (5.5+0.707*\l,-0.5-0.707*\l)  to (5.5+0.707*\l,-0.5+0.707*\l) to  (5.5-0.707*\l,-0.5+0.707*\l);
\draw [c3!60,very thick] (5.5-0.707*\l,-0.5+0.707*\l) to (5.5+0.707*\l,-0.5-0.707*\l);
\draw [c3!60,very thick] (5.5-0.707*\l,-0.5-0.707*\l) to (5.5+0.707*\l,-0.5+0.707*\l);
\draw [draw opacity=0,outer color=c3!60,inner color=c1](5.5-0.707*\l,-0.5+0.707*\l) circle (\r);	
\draw [draw opacity=0,outer color=c3!60,inner color=c1](5.5-0.707*\l,-0.5-0.707*\l) circle (\r);		
\draw [draw opacity=0,outer color=c3!60,inner color=c1](5.5+0.707*\l,-0.5-0.707*\l) circle (\r);	
\draw [draw opacity=0,outer color=c3!60,inner color=c1](5.5+0.707*\l,-0.5+0.707*\l) circle (\r);	
%     \node   at (5.5-0.707*\l,-0.5+0.707*\l) {\sffamily\tiny Q6};
%    \node  at (5.5-0.707*\l,-0.5-0.707*\l) {\sffamily\tiny Q7};		
%    \node   at(5.5+0.707*\l,-0.5-0.707*\l) {\sffamily\tiny Q8};
%     \node  at (5.5+0.707*\l,-0.5+0.707*\l) {\sffamily\tiny Q9};	
     
  \draw [latex-latex, black] (0+3*0.275,0+2*0.275) to (0+3*0.725,0+2*0.725);
  \draw [latex-latex, black] (0+5.5*0.175,0-0.5*0.175) to (0+5.5*0.825,0-0.5*0.825);
  \draw [latex-latex, black] (3+2.5*0.275,2-2.5*0.275) to (3+2.5*0.725,2-2.5*0.725);
  \draw [latex-, black] (3+3*0.325,2+0.5*0.325) to (3+3*0.7,2+0.5*0.7); 
  \draw[dashed, black] (3+3*0.725,2+0.5*0.725)to (3+3*0.9,2+0.5*0.9);
   \draw [latex-, black] (5.5+0.5*0.325,-0.5+3*0.325) to (5.5+0.5*0.7,-0.5+3*0.7); 
   \draw[dashed, black] (5.5+0.5*0.7,-0.5+3*0.7) to (5.5+0.5*0.9,-0.5+3*0.9);
   
   \node [below, black] at (0,-1) {\footnotesize\textsc{Subsystem 1}};
   \node [below, black] at (3,1) {\footnotesize\textsc{Subsystem 2}};
    \node [below, black] at (5.5,-1.5) {\footnotesize\textsc{Subsystem 3}};      
\end{tikzpicture}
\caption{Schematic of   a block architecture. Physically,  it is appropriate to assume that  the longer  range couplings  are negligibly small. This forms a sound   basis for dividing the entire architecture into blocks. Each block represents a subsystem of coupled qubits, and  the single arrows in  between    represent inter-subsystem couplings.  Subsystem-based QOC   performs pulse optimization on the subsystems,   it   has a much reduced problem scaling.    }
\label{subsystems}
\end{figure}

It  is worth pointing out that the     feasibility of the approach crucially relies on      assuming that  the inter-subsystem couplings can be treated  as  perturbation terms. In principle, the validity of the assumption is relevant to how the full system is divided. Yet from a practical aspect of view, this assumption is  reasonable because a realistic physical architecture    normally has limited qubit connectivity (e.g., nearest-neighbor couplings only)  like that   shown in Fig. \ref{subsystems}. The couplings between distant qubits are more likely to be weaker than those between adjacent qubits. This fact makes it possible to keep   the subsystems   small, and thence the problem size would grow only polynomially.
To demonstrate the potential of  the subsystem-based QOC method, I optimize pulses for elementary quantum gates on an NMR 12-qubit  multiple coupled system. The results show that, due to the avoidance of  simulations on the full system, the   method is able to offer orders of magnitude speedup in finding pulses with around $ 99\%$ fidelities.

\emph{Subsystem-based QOC.---}%For a complex matrix $V$, $\left\| \cdot\right\|$ stands for the Frobenius norm $\left\| V \right\| = \sqrt{\operatorname{Tr}(V V^\dag)}$. Let $I_d$ denote the $d$-dimensional identity operator.
To start, I describe the basic problem setting. Consider an $n$-qubit quantum  system $S$ described by a  system Hamiltonian $H_S$. The task is to implement a desired unitary operation $\overline U$. Our framework  restricts considerations to   local quantum gates that operate on only a few qubits at a time. More general quantum operations can be decomposed into an array of simpler local quantum gates. We divide the system into $s$ disjoint subsystems $S =  S_1 \cup \cdots \cup S_s$. The choice of the division is governed by  two considerations: (i) there are only a few comparatively small couplings between the subsystems, and (ii) the qubits that      $\overline U$ operates on belong to the same subsystem, so that  one can write $\overline U = \overline U_{S_1} \otimes \cdots \otimes \overline U_{S_s}$.  With this subsystem division, the system Hamiltonian can now be written as
\begin{equation}
	H_S = H_0 + H_1 = \sum_{k=1}^s H_{S_k} + \sum_{k < j}^s H_{S_k S_j},
	 \end{equation}
where $H_{S_k}$ is the internal Hamiltonian of subsystem $S_k$, and $H_{S_k S_j}$ is the interaction Hamiltonian between $S_k$ and $S_j$. We denote the Hilbert space of the $k$th subsystem   $S_k$ as  $\mathcal{H}_{S_k}$, so $\left\{ H_{S_k} \right\}$ are operators on  $ \mathcal{H}_k$, and $\left\{ H_{S_k S_j} \right\}$ are operators on $\mathcal{H}_{S_k} \otimes  \mathcal{H}_{S_j} $. To steer the system towards the desired operation, we   add coherent controls  based on   externally applied control fields  $u(t)$ ($0\le t \le T$). 
Often,   the control fields are coupled with the individual qubits, hence we shall assume that the control Hamiltonian takes   the form $H_C(t) = \sum_k  H_{C_k}(t)$, in which $H_{C_k}(t)$ is the control Hamiltonian of subsystem $S_k$.   
The basic approach to the   optimal control problem for the whole system is to search for a pulse shape $u(t)$ to
\begin{align}
\label{QOC}
	\max \quad  &  F =   \left| \operatorname{Tr}(U(T) \overline U^\dag) \right|^2/d^2,  \\
    \text{s.t.} \quad & \dot U(t) = -i[H_S +H_C(t)]U(t). \nonumber
\end{align}
where $F$ is the fidelity function, $d=\dim \mathcal{H}$, and $U$ starts from the identity operation $U(0) = \bm{1}_d$. 

The basic idea of our approach is to consider a related robust quantum control problem. That is, we solve the optimal control problem for the intra-subsystem Hamiltonian $H_0$, and at the meantime require   the control solution be robust to the inter-subsystem Hamiltonian $H_1$.   For now, we denote $U(t)$ as the propagator generated by $H_0 + H_C(t)$, and view $H_1$ as a variation of the generator. Let  $\mathcal{D}_{U(T)}(H_{S_k S_j})$ denote the directional derivative of the propagator $U(T)$ with respect to the variation in $H_0 + H_C(t)$ in the direction $H_{S_k S_j}$, as given by
\begin{align}
	\mathcal{D}_{U(T)}(H_{S_k S_j}) & = \lim_{\epsilon \to 0} \frac{U(T,\epsilon) - U(T)}{\epsilon}    \nonumber \\
&	=   \left.\frac{d}{d\epsilon}\right|_{\epsilon =0} \mathcal{T} e^{  -i \int_0^T dt \left[  H_0 + H_C(t) + \epsilon H_{S_k S_j} \right]   } \nonumber
\end{align}
where $\mathcal{T}$ is the   time-ordered operator.    Then,   robustness in $H_1$ can be achieved via maximizing the fidelity function $f =   \left| \operatorname{Tr}(U(T) \overline U^\dag) \right|^2/d^2$ and minimizing the norm of every directional derivative $f_{S_k S_j} = \left\| \mathcal{D}_{U(T)}(H_{S_k S_j}) \right\|^2/d^2$. Now, a critical observation is that it suffices to perform computations on just the  subsystems. First, notice that for any time $t$, $U(t)$ can be factorized as $U(t) = U_{S_1}(t) \otimes \cdots \otimes U_{S_s}(t)$, where $U_{S_k}(t)$ is the propagator of subsystem $S_k$ generated by $H_{S_k} + H_{C_k}(t)$, therefore  $f$ equals to the product of the fidelities of the subsystems $f  = \prod_{k=1}^s f_{S_k}$. As to the directional derivatives, there is   
\begin{align}
	f_{S_k S_j} & =  \left\| -i U(T) \int_0^T dt U^\dag(t) H_{S_k S_j} U(t) \right\|^2 /d^2  \nonumber \\
	& =   \left\| \mathcal{D}_{U_{S_k}(T) \otimes  U_{S_j}(T)}(H_{S_k S_j}) \right\|^2 /d^2_{\mathcal{H}_{S_k} \otimes \mathcal{H}_{S_j}}.  \nonumber
\end{align}
Here,   the   first line can be obtained via the Dyson series \cite{PhysRev.75.486,HPZC19}. Therefore, $f_{S_k S_j}$ can be evaluated simply on the pair of subsystems $S_k$ and $S_j$. More details about the derivation can be found in the Supplementary Material \cite{SupplementalMaterial}.

\begin{figure*} 
\begin{center}
\includegraphics[width=\linewidth]{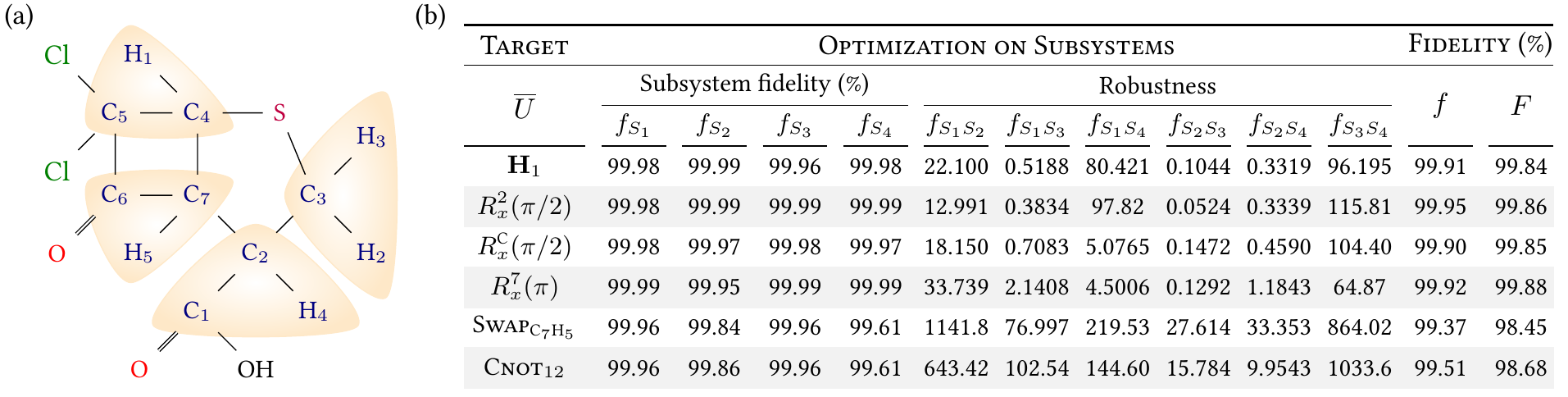}
\end{center}
\caption{(a) Molecular structure  of per-$^{13}$C-labeled dichlorocyclobutanone. It contains
7 labeled carbon nuclei and 5 proton nuclei and hence
forms a 12-qubit coupled system. The spins are coupled via through-bond interaction, so  those spins that are not connected via chemical bonds would have smaller coupling strengths. (b) Subsystem-based pulse optimization. The method is tested for    a number of different kind of local target operations. Here, $\mathbf{H}$ is the Hadamard gate, $R^k_x(\theta)$ denotes a rotational gate that rotates qubit $k$ about the $x$ axis with rotational angle $\theta$.  As defined in the main text, $f$ and $F$ is the pulse fidelity in terms of subsystem Hamiltonian $H_0$ and full system Hamiltonian $H_S$, respectively.}	
\label{result}
\end{figure*}

Summarizing the above analysis, the robust control problem  is   essentially   a    multi-objective optimization  problem on subsystems
\begin{align}
\label{SRQOC}
	\max \quad  &  \Phi =   \prod_{k=1}^s f_{S_k} - \sum_{k<j}^s \lambda_{kj} f_{S_k S_j},  \\
    \text{s.t.} \quad & \dot U_{S_k }(t)  =   -i  [H_{S_k}    +  H_{C_k}(t)   ] U_{S_k}(t). \nonumber 
\end{align}
Here,  the fitness function $\Phi$   is chosen as a linear combination of the objective functions, with $\lambda_{kj}$ being the positive weight coefficients associated to individual objectives. The  weights will be   tuned  within an iterative algorithmic procedure. Since $f_{S_k S_j} \ge 0$, there must be $\Phi \le 1$, so maximizing $\Phi$ close to 1 will imply a   solution to the robust control problem Eq. (\ref{SRQOC}), which also provides a solution to the original problem Eq. (\ref{QOC}). 
This constitutes the central result of this work.

Now let us   turn our attention to the solution method of the problem Eq. (\ref{SRQOC}).
There have been developed a variety of quantum optimal control algorithms. In particular, GRAPE, a gradient-based technique, is widely   used for its   good convergence speed and high numerical accuracy. In GRAPE, the optimization   consists of iteratively executing two major steps. First the gradient of the fitness function with respect to the control parameters is computed, and then one determines an appropriate step length along a gradient-related direction. The algorithm continues until, an optimal solution is found or a termination criterion is satisfied. For the present problem, suppose the pulse $u(t)$ is discretized into $M$ slices of equal length $\tau = T/M$ and that   the $m$th   slice being of constant magnitude $u[m]$, then the key step is to   compute the gradient   $\partial \Phi/ \partial u[m]$. To this end, I  follow the methodology developed by Ref. \cite{HPZC19},   which   extends the Shr{\"o}dinger equation to a so-called Van Loan block matrix differential equation by which means the same gradient-based algorithm can  apply. Concretely, define a set of operators $\left\{L_{S_k S_j}: 1\le k < j \le s \right\}$ and $\left\{ L_{C_k}(t): k =1,\ldots, s \right\}$ by 
\begin{align}
	L_{S_k S_j} & = \left[\begin{matrix}
H_{S_k} + H_{S_j}    &  H_{S_k S_j} \\
0  & H_{S_k} + H_{S_j}   \\
\end{matrix}
\right], \nonumber \\
L_{C_k }(t) & = \left[\begin{matrix}
H_{C_k}(t)     &  0 \\
0  & H_{C_k}(t)   \\
\end{matrix}
\right]. \nonumber
\end{align}
They are referred to as Van Loan generators \cite{VanLoan78,CJP08}. Construct the Van Loan differential equation as 
\[ \dot V_{S_k S_j}(t)  =   -i  [L_{S_k S_j}    +  L_{C_k}(t)  + L_{C_j}(t) ] V_{S_k S_j}(t), \]  
then there is for   $V_{S_k S_j}(t)$  the following expression
\begin{equation}
	V_{S_k S_j}(t) = \left[\begin{matrix}
U_{S_k S_j}(t)     &  \mathcal{D}_{U_{S_k S_j}(t)} (H_{S_k S_j}) \\
0  & U_{S_k S_j}(t)   \\
\end{matrix}
\right]. \nonumber
\end{equation}  
Based on the above formula,  the problem Eq. (\ref{SRQOC}) is therefore essentially to optimize the   following fitness function  
\begin{equation}
\label{SVanLoanQOC}
	\Phi =   \prod_{k=1}^s f_{S_k} - \sum_{k<j}^s \lambda_{kj} \left\| V^{(1,2)}_{S_k S_j}(T) \right\|^2/d^2_{\mathcal{H}_{S_k} \otimes \mathcal{H}_{S_j}},
\end{equation} 
in which $V^{(1,2)}_{S_k S_j}(T)$ is the $(1,2)$ block sub-matrix of $V_{S_k S_j}(T)$. As observed by Ref. \cite{HPZC19}, since the Van Loan equation has   the same general form as the Schr{\"o}dinger equation,  which implies that Eq. (\ref{SVanLoanQOC}) still belongs to the class of bilinear control theory problems \cite{Elliott09}, thus evaluation of $\partial \Phi/ \partial u[m]$ can be done in the same way as how  gradients are estimated in GRAPE. More details about the derivations and the algorithmic procedure can be found in the Supplementary Material \cite{SupplementalMaterial}. 
%An important technicality is the tuning of the parameters $\lambda_{kj}$.

%If the system consists of $n$ qubits. We divide it into $S$ parts, so on average the dimension of the subsystem is $2^{n/S}$. If we consider robustness up to order $m$, then the subsystem-based GRAPE should handle matrices of dimension $m 2^{n/S}$.

\emph{Optimal control on a 12-qubit NMR system.---}The utility of the developed scheme can be demonstrated with a 12-qubit NMR  molecule  ($^{13}$C-labeled dichlorocyclobutanone)  \cite{Dawei17,PhysRevLett.123.030502}. This is to date the largest spin  quantum information processor with full controllability;  see Supplemental Materials for its Hamiltonian parameters   \cite{SupplementalMaterial}. Running GRAPE for as large as a 12-qubit system  involves  computing of $2^{12}$-dimensional  matrix exponentials and matrix multiplications as well, which requires a  substantial amount of memory   and time cost. The subsystem-based approach circumvents this difficulty.  To see this, I divide the whole system into four subsystems, $S_1 = \left\{ \text{C}_1,\text{C}_2,\text{H}_4 \right\}$, $S_2 = \left\{ \text{C}_3,\text{H}_2,\text{H}_3 \right\}$, $S_2 = \left\{ \text{C}_4,\text{C}_5,\text{H}_1 \right\}$, and $S_4 = \left\{ \text{C}_6,\text{C}_7,\text{H}_5 \right\}$, each consisting of three spins; see Fig. \ref{result}(a).    Instead of searching on the whole system, I  perform pulse optimization     on these    subsystems, and consider pulse robustness to the inter-subsystem couplings $\left\{ H_{S_k S_j}\right\}$ ($1\le k < j \le 4$). In this way, the optimization process involves only $2 \times 2^6$-dimensional matrix operations.     The results are shown in Fig. \ref{result}(b);   more  data are given in Supplementary Material \cite{SupplementalMaterial}. 
From the simulations, several features can be identified. First, for a particular target operation, larger inter-subsystem couplings would result in larger perturbations to the ideal operation. Second, if a pulse has longer time length, the  inter-subsystem couplings will have larger effects to the controlled evolution, and it will be harder to enhance the robustnes. On the whole,     for  the optimized pulses, the decreases from       their full system fidelities    to their corresponding subsystem fidelities are smaller than   1\%. This confirms the capability of the subsystem-based algorithm in producing high-fidelity controls.

\emph{Discussions and Outlook.---}The key insight behind the subsystem-based   QOC method   is the recognization that    a practical  physical system should   have   limited range couplings, so subsystem-based strategies can apply. Although all-to-all connected qubit graph  may facilitate implementation of certain quantum tasks like quantum entanglement generation and has been experimentally realized in, e.g.,  small trapped ion systems \cite{Monroe16}, such type of architecture is unlikely to scale up to large number of qubits. For a general   setting, depending on the concrete qubit connectivity graph and the particular optimal control problem of interest, one can view  some of the small couplings as perturbations or even simply  ignore them   to  obtain  a subsystem-based problem simplification.     Gradient-based algorithms have been employed in solving the reduced optimization problem and show  good applicability.   For further studies,  it will  be interesting to consider using gradient-free algorithms    as  they have the potential in producing globally optimal solutions.

Precise coherent control of quantum systems is an essential prerequisite for various quantum technologies. However,  noisy intermediate-scale quantum  devices \cite{Preskill2018quantumcomputingin} with   tens to   hundreds of qubits that are about to appear in the near-term already push   most QOC methods to their limit, thus posing an urgent demand for   research of scalable QOC.    In developing a scalable QOC method, the fundamental objective is to avoid as many   full quantum system simulations as possible, so as to keep the problem scale tractable. This work achieves this goal, and the presented   algorithm  could become a useful tool to be incorporated into  a  control software layer to enhance the performance of quantum hardware for tasks like computing and simulation  \cite{PhysRevX.8.021059}.
It is thus believed that the methodology   here will promote studies of quantum technologies on future large   quantum systems.

\emph{Acknowledgments}. J. L. is supported by the National Natural Science Foundation of China (Grants No. 11605005, No. 11875159, No. 11975117, and No. U1801661), Science, Technology and Innovation Commission of Shenzhen Municipality (Grants No. ZDSYS20170303165926217, No. JCYJ20170412152620376, and
No. JCYJ20180302174036418)), Guangdong Innovative
and Entrepreneurial Research Team Program (Grant No. 2016ZT06D348).

%\bibliography{manuscript} 


\newpage
\appendix
\onecolumngrid

\vspace{100pt}
\begin{center}
{\large\bfseries Supplementary Material}	
\end{center}

\section{Theory}
In the main text of our manuscript, we stated that the central result of our work is to show that one can solve the quantum optimal control problem
\begin{align}
\label{QOC}
	\max \quad  &  F =   \left| \operatorname{Tr}(U(T) \overline U^\dag) \right|^2/d^2,  \\
    \text{s.t.} \quad & \dot U(t) = -i[H_S +H_C(t)]U(t). \nonumber
\end{align}
by solving a  multi-objective optimization problem on the subsystems $\left\{S_k\right\}$:
\begin{align}
\label{SRQOC}
\max   \quad  &   \Phi  =   f - \sum_{k<j}^s \lambda_{kj} f_{S_k S_j}   = \prod_{k=1}^s \left| \operatorname{Tr}(U_{S_k}(T) \overline U_{S_k}^\dag) \right|^2/d^2_{\mathcal{H}_{S_k}} - \sum_{k<j}^s \lambda_{kj} \left| \mathcal{D}_{U_{S_k S_j}(T)}(H_{S_k S_j}) \right\|^2/d^2,   \\
\text{s.t.} \quad  &  \dot U_{S_k}(t) = -i  [H_{S_k}  + H_{C_k}(t) ] U_{S_k}(t).   \nonumber 
\end{align}
which is further converted to optimal control problem under constraints of block matrix Van Loan differential equations:
\begin{alignat}{2}
\label{SVanLoanQOC}
\max & \quad  &   \Phi & = \prod_{k=1}^s f_{S_k} - \sum_{k<j}^s \lambda_{kj} f_{S_k S_j}    = \prod_{k=1}^s \left| \operatorname{Tr}(U_{S_k}(T) \overline U_{S_k}^\dag) \right|^2/d^2_{\mathcal{H}_{S_k}} - \sum_{k<j}^s \lambda_{kj} \left\| V^{(1,2)}_{S_k S_j}(T) \right\|^2/d^2_{\mathcal{H}_{S_k} \otimes \mathcal{H}_{S_j}},   \\
\text{s.t.} &   & \dot U_{S_k }(t)
    & =   -i  [H_{S_k}    +  H_{C_k}(t)   ] U_{S_k}(t), \nonumber  \\
   &   & \dot V_{S_k S_j}(t)
    & =    -i  [L_{S_k S_j}    +  L_{C_k}(t)  + L_{C_j}(t) ] V_{S_k S_j}(t) \nonumber  \\
 & &  & = -i\left[\begin{matrix}
H_{S_k} + H_{S_j} +  H_{C_k}(t) + H_{C_j}(t)   &  H_{S_k S_j} \\
0  & H_{S_k} + H_{S_j} +  H_{C_k}(t) + H_{C_j}(t)  \\
\end{matrix}
\right]V_{S_k S_j}(t). \nonumber   
\end{alignat}
This section is devoted to explaining the following contents, so as to support the validity and applicability arguments that we have described for our method   in the main text: 
\begin{enumerate}
	\item The directional derivative $\mathcal{D}_{U(T)}(H_{S_k S_j})$ has  the following analytic expression
	\begin{equation}
		\mathcal{D}_{U(T)}(H_{S_k S_j}) = -i U(T) \int_0^T dt U^\dag(T)H_{S_k S_j}U(T).
	\end{equation}
	\item Van Loan integrals involving time-ordered matrix exponentials.
	\item Computational methods for Van Loan differential equations.
\end{enumerate}
We should remark that the analysis in the below   largely follows the approach developed in Ref. \cite{HPZC19}, but it is beneficial for us to   present them here in detail for reader's convenience.

\subsection{Directional Derivatives From Dyson Series Expansion}
Consider  the perturbed evolution $U(t,\epsilon)$ when there is  a variation in the generator $H_0 + H_C(t)$ in the direction $H_{S_k S_j}$
\[ \dot U(t,\epsilon) = -i[H_S +H_C(t) + \epsilon H_{S_k S_j}]U(t,\epsilon). \]
Define the transformation of representation as $\widetilde U(t,\epsilon) = U^\dag(t) U(t,\epsilon)$, then
\begin{align}
	\frac{\partial \widetilde U(t,\epsilon) }{\partial t} &  = \dot U^\dag(t) U(t,\epsilon) +   U^\dag(t) \dot U(t,\epsilon) \nonumber \\
	& = i U^\dag(t) [H_S +H_C(t)]   U(t,\epsilon) -i U^\dag(t)  [H_S +H_C(t) + \epsilon H_{S_k S_j}]U(t,\epsilon) \nonumber \\
	& = -i U^\dag(t) \epsilon H_{S_k S_j} U(t,\epsilon) \nonumber \\
	& = -i U^\dag(t) \epsilon H_{S_k S_j} U(t) \widetilde U(t,\epsilon). \nonumber
\end{align}
The formal solution for $\widetilde U(t,\epsilon)$ is then given as
\[\widetilde U(T,\epsilon) = -i \int_0^T dt U^\dag(t) \epsilon H_{S_k S_j} U(t)  \]
Turn back to the original representation, one gets
\[ U(T,\epsilon) = -i U(T) \int_0^T dt U^\dag(t) \epsilon H_{S_k S_j} U(t). \]
Now expand the right hand side of the above formula via the Dyson series \cite{PhysRev.75.486}
\[ U(T,\epsilon) =   U(T) \left(\mathbb{I}_{2^n} -i \epsilon \int_0^T dt U^\dag(t)   H_{S_k S_j} U(t) - \epsilon^2 \int_0^T dt_1 \int_0^{t_1} dt_2 U^\dag(t_1)   H_{S_k S_j} U(t_1) U^\dag(t_2)   H_{S_k S_j} U(t_2) + \cdots \right). \]
from this there is
\begin{equation}
	\mathcal{D}_{U(T)}(H_{S_k S_j})   = \lim_{\epsilon \to 0} \frac{U(T,\epsilon) - U(T)}{\epsilon} =  -i  U(T) \int_0^T dt U^\dag(t)   H_{S_k S_j} U(t).  
\end{equation}
This gives an applicable formula for evaluating the directional derivatives.

\subsection{Block Matrix Van Loan Differential Equation Framework}
Van Loan  originally  observed   the following expression \cite{VanLoan78}:
\[ \exp\left[ \left( 
\begin{matrix}
  A & B \\
 0 & A \\
\end{matrix}
 \right) t \right] =  \left( 
\begin{matrix}
  e^{At} & e^{At} \int_0^t dt_1 e^{-At_1} B e^{At_1} \\
 0 & e^{At} \\
\end{matrix}
 \right).
 \]
Ref. \cite{HPZC19} extended Van Loan's formula to  integrals involving time-ordered exponentials, applying which to the present problem will give
\begin{align}
{} & \mathcal{T} \exp\left[ -i \int_0^T dt \left(  \begin{matrix}
H_{S_k} + H_{S_j} +  H_{C_k}(t) + H_{C_j}(t)   &  H_{S_k S_j} \\
0  & H_{S_k} + H_{S_j} +  H_{C_k}(t) + H_{C_j}(t)  \\
\end{matrix}
\right)   \right] \nonumber \\
= {} & \left( 
\begin{matrix}
  U_{S_k S_j}(T) & -i  U_{S_k S_j}(T) {\displaystyle \int_0^T dt U_{S_k S_j}^\dag(t)   H_{S_k S_j} U_{S_k S_j}(t) }\\
 0 & U_{S_k S_j}(T) \\
\end{matrix}
 \right)  \nonumber \\
= {} & \left( 
\begin{matrix}
  U_{S_k S_j}(T) & \mathcal{D}_{U_{S_k S_j}(T)}(H_{S_k S_j}) \\
 0 & U_{S_k S_j}(T) \\
\end{matrix}
 \right).	
\end{align} 
This implies that we could compute  $\mathcal{D}_{U_{S_k S_j}(T)}(H_{S_k S_j})$ through solving the differential equation
\[ \dot V_{S_k S_j}(t) = -i\left[\begin{matrix}
H_{S_k} + H_{S_j} +  H_{C_k}(t) + H_{C_j}(t)   &  H_{S_k S_j} \\
0  & H_{S_k} + H_{S_j} +  H_{C_k}(t) + H_{C_j}(t) \end{matrix} \right] V_{S_k S_j}(t), \]
with $V_{S_k S_j}(0) = \mathbb{I}_{d_{\mathcal{H}_{S_k} \otimes \mathcal{H}_{S_j}} } \oplus \mathbb{I}_{d_{\mathcal{H}_{S_k} \otimes \mathcal{H}_{S_j}} }$. The result for $\mathcal{D}_{U_{S_k S_j}(T)}(H_{S_k S_j})$ can then be extracted from the $(1,2)$-th block sub-matrix of $V_{S_k S_j}(T)$. The idea can be generalized to nested integrals of arbitrary order, so that one can   compute directional derivatives of higher order when necessary. We refer the readers to Ref. \cite{HPZC19} for more descriptions of the theory.

\subsection{Computational Methods}
The Gradient Ascent Pulse Engineering (GRAPE) algorithm
proposed in Ref. \cite{Khaneja05} is a standard numerical method for solving  quantum optimal control problems. GRAPE works as follows
\begin{enumerate}
	\item[(1)] start from an initial pulse guess $u[m]$, $m=1,\ldots,M$;
	\item[(2)] compute the gradient of the fitness function with respect to the pulse parameters $\partial \Phi / \partial u[m]$;
	\item[(3)] determine a gradient-related search direction $p[m]$, $m=1,\ldots,M$;
	\item[(4)] determine an appropriate step length $l$ along the search direction $p$;
	\item[(5)] update the pulse parameters: $u[m] \leftarrow u[m] + l \times p[m]$;
	\item[(6)] if $\Phi$ is sufficiently high, terminate; else, go to step 2.
\end{enumerate}
For our present   problem Eq. (\ref{SVanLoanQOC}), we now describe how to evaluate the gradient $\partial \Phi / \partial u[m]$ in the following.

Suppose the pulse has total time length $T$, and is    discretized into $M$ pieces with each piece   of duration $\tau = T/M$. Denote the propagators $U_{S_k}[m]$ and $V_{S_k S_j}[m]$ of the $m$th piece as
\begin{align}
	U_{S_k}[m] & = \exp\left( -i [H_{S_k} + H_{C_k}(t)] \tau \right), \nonumber \\
	V_{S_k S_j}[m] & = \exp\left( -i [L_{S_k S_j} + L_{C_k}(t)+ L_{C_j}(t)] \tau \right).  \nonumber
\end{align}
Then 
\begin{equation}
	\frac{\partial \Phi}{\partial u[m]} =  \sum_{j=1}^s  \left(\frac{\partial f_{S_j}}{\partial u[m]} \prod_{k\ne j}^s f_{S_k} \right)  - \sum_{k<j}^s \lambda_{kj} \frac{\partial f_{S_k S_j}}{\partial u[m]},
\end{equation}
in which
\begin{align}
	\frac{\partial f_{S_k}}{\partial u[m]} & = 2 \operatorname{Re} \left\{ \operatorname{Tr}\left(\frac{\partial U_{S_k}(T)}{\partial u[m]}  \overline U_{S_k}^\dag \right) \left(\operatorname{Tr}(U_{S_k}(T)  \overline U_{S_k}^\dag \right)^*\right\}  /d^2_{\mathcal{H}_{S_k}}  \nonumber \\
	& = 2 \operatorname{Re} \left\{ \operatorname{Tr}\left(-i \tau U_{S_k}[M] \cdots \frac{\partial H_{C_k}[m]}{\partial u[m]} U_{S_k}[m] \cdots U_{S_k}[1] \overline U_{S_k}^\dag \right) \left(\operatorname{Tr}(U_{S_k}(T)  \overline U_{S_k}^\dag \right)^*\right\}  /d^2_{\mathcal{H}_{S_k}}, \nonumber
\end{align}
and
\begin{align}
	\frac{\partial f_{S_k S_j}}{\partial u[m]} & =  2 \operatorname{Re} \left\{ \operatorname{Tr} \left[ \frac{\partial V^{(1,2)}_{S_k S_j}(T)}{\partial u[m]}  \left( V^{(1,2)}_{S_k S_j}(T) \right)^\dag \right] \right\}  /d^2_{\mathcal{H}_{S_k} \otimes \mathcal{H}_{S_j}} \nonumber \\
	& =  2 \operatorname{Re} \left\{ \operatorname{Tr} \left[ \left( -i \tau  V_{S_k S_j}[M] \cdots \frac{\partial (L_{C_k}[m] + L_{C_j}[m])}{\partial u[m]}  V_{S_k S_j}[m]   \cdots V_{S_k S_j}[1] \right)^{(1,2)}  \left( V^{(1,2)}_{S_k S_j}(T) \right)^\dag \right] \right\}  /d^2_{\mathcal{H}_{S_k} \otimes \mathcal{H}_{S_j}}. \nonumber
\end{align}

%\subsection{Estimating Overall Fidelity $F$}  
%We make an estimate of the overall fidelity via series expansion. Denote $\epsilon = \left\{ \epsilon_{kj}: 1\le k < j \le s \right\}$
%\begin{align}
%	f(\epsilon) & = \left| \operatorname{Tr}(U(T,\epsilon) \overline U_{S_k}^\dag) \right|^2/d^2  \nonumber  \\
%	& \approx \left| \operatorname{Tr}\left\{ \left[ U(T) + \sum_{k<j}^s \epsilon_{kj} \mathcal{D}_{U(T)}(H_{S_k S_j})   \right]  \overline U_{S_k}^\dag \right\} \right|^2/d^2
%\end{align}

\begin{figure}[t]
\begin{center}
\begin{tikzpicture}[scale=0.017]

  \node at (-110,90) {  (a)};
  \node at (-110,-180) {  (d)};
   
  \node (C5) [blue!50!black] at (-30,30) { \small C$_5$};
  \node (C4) [blue!50!black] at (30,30) {\small C$_4$};
  \node (C6) [blue!50!black] at (-30,-30) {\small C$_6$};
  \node (C7) [blue!50!black] at (30,-30) {\small C$_7$};
  \draw (C7) to (C4) to (C5) to (C6) to (C7);
  
  \node (C2) [blue!50!black] at (30+30*1.414,-30-30*1.414) {\small C$_2$};
  \draw (C7) to (C2);
  \node (C1) [blue!50!black] at (30,-30-60*1.414) {\small C$_1$};
  \draw (C1) to (C2);
  \node (C3) [blue!50!black] at (30+60*1.414,-30) {\small C$_3$};
  \draw (C3) to (C2);
  
  \node (H1) [blue!50!black] at (30-30*1.414,30+30*1.414) {\small H$_1$};
  \draw (C4) to (H1);
  \node (H2) [blue!50!black] at (30+90*1.414,-30+30*1.414) {\small H$_3$};
  \draw (C3) to (H2);
  \node (H3) [blue!50!black] at (30+90*1.414,-30-30*1.414) {\small H$_2$};
  \draw (C3) to (H3);
  \node (H4) [blue!50!black] at (30+60*1.414,-30-60*1.414) {\small H$_4$};
  \draw (C2) to (H4);
  \node (H5) [blue!50!black] at (30-30*1.414,-30-30*1.414) {\small H$_5$};
  \draw (C7) to (H5);

  \node (OH) at (30+30*1.414,-30-90*1.414) {\small OH};
  \draw (C1) to (OH);
  \node (O1) [red] at (30-30*1.414,-30-90*1.414) {\small O};
  \draw [double] (C1) to (O1);
  \node (O2) [red] at (-30-30*1.414,-30-30*1.414) {\small O};
  \draw [double] (C6) to (O2);
  \node (Cl1) [green!50!black] at (-30-30*1.414,30-30*1.414) {\small Cl};
  \draw (C5) to (Cl1);
  \node (Cl2) [green!50!black] at (-30-30*1.414,30+30*1.414) {\small Cl};
  \draw (C5) to (Cl2);
  \node (S) [purple] at (90,30) {\small S};
  \draw (C4) to (S) to (C3);

\end{tikzpicture}\hspace{0.02\linewidth}
\begin{tikzpicture}[scale=0.017]

  \node at (-110,90) {   (b)};
  \node at (-110,-180) { };
 \draw [c1,outer color=c1,inner color=white,xshift=-80] plot [smooth cycle] coordinates {(-15.2598,97.2582)(-46.3227,48.9359)(-46.3227,-48.9359)(-15.2598,-97.2582)(48.7796,-46.5023)(48.7796,46.5023)};

\draw [c1,outer color=c1,inner color=white,xshift=-80,yshift=80] plot [smooth cycle] coordinates {(8.9085,-128.2619)(118.3755,-139.5887)(182.0087,-75.9555)(170.6819,33.5115)(119.7429,-5.4855)(47.9055,-77.3229)};

%\draw [c1,outer color=c1,inner color=white] plot [smooth cycle] coordinates {(8.79,-125.445)(114.84,-125.445)(167.865,-72.42)(167.865,33.63)};
 
  \node (C5) [blue!50!black] at (-30,30) { \small C$_5$};
  \node (C4) [blue!50!black] at (30,30) {\small C$_4$};
  \node (C6) [blue!50!black] at (-30,-30) {\small C$_6$};
  \node (C7) [blue!50!black] at (30,-30) {\small C$_7$};
  \draw (C7) to (C4) to (C5) to (C6) to (C7);
  
  \node (C2) [blue!50!black] at (30+30*1.414,-30-30*1.414) {\small C$_2$};
  \draw (C7) to (C2);
  \node (C1) [blue!50!black] at (30,-30-60*1.414) {\small C$_1$};
  \draw (C1) to (C2);
  \node (C3) [blue!50!black] at (30+60*1.414,-30) {\small C$_3$};
  \draw (C3) to (C2);
  
  \node (H1) [blue!50!black] at (30-30*1.414,30+30*1.414) {\small H$_1$};
  \draw (C4) to (H1);
  \node (H2) [blue!50!black] at (30+90*1.414,-30+30*1.414) {\small H$_3$};
  \draw (C3) to (H2);
  \node (H3) [blue!50!black] at (30+90*1.414,-30-30*1.414) {\small H$_2$};
  \draw (C3) to (H3);
  \node (H4) [blue!50!black] at (30+60*1.414,-30-60*1.414) {\small H$_4$};
  \draw (C2) to (H4);
  \node (H5) [blue!50!black] at (30-30*1.414,-30-30*1.414) {\small H$_5$};
  \draw (C7) to (H5);

  \node (OH) at (30+30*1.414,-30-90*1.414) {\small OH};
  \draw (C1) to (OH);
  \node (O1) [red] at (30-30*1.414,-30-90*1.414) {\small O};
  \draw [double] (C1) to (O1);
  \node (O2) [red] at (-30-30*1.414,-30-30*1.414) {\small O};
  \draw [double] (C6) to (O2);
  \node (Cl1) [green!50!black] at (-30-30*1.414,30-30*1.414) {\small Cl};
  \draw (C5) to (Cl1);
  \node (Cl2) [green!50!black] at (-30-30*1.414,30+30*1.414) {\small Cl};
  \draw (C5) to (Cl2);
  \node (S) [purple] at (90,30) {\small S};
  \draw (C4) to (S) to (C3);

\end{tikzpicture}
\hspace{0.02\linewidth}
\begin{tikzpicture}[scale=0.017]

  \node at (-110,90) {   (c)};
  \node at (-110,-180) { };

\draw [c1,outer color=c1,inner color=white] plot [smooth cycle] coordinates {(-18.63,93.63)(-49.395,19.395)(55.605,19.395)};
\draw [c1,outer color=c1,inner color=white] plot [smooth cycle] coordinates {(-49.395,-19.395)(55.605,-19.395)(-18.63,-93.63)};
\draw [c1,outer color=c1,inner color=white] plot [smooth cycle] coordinates {(-1.815,-125.445)(72.42,-51.21)(146.655,-125.445)};
\draw [c1,outer color=c1,inner color=white] plot [smooth cycle] coordinates {(93.63,-30)(167.865,-104.235)(167.865,44.235)};

  \node (C5) [blue!50!black] at (-30,30) { \small C$_5$};
  \node (C4) [blue!50!black] at (30,30) {\small C$_4$};
  \node (C6) [blue!50!black] at (-30,-30) {\small C$_6$};
  \node (C7) [blue!50!black] at (30,-30) {\small C$_7$};
  \draw (C7) to (C4) to (C5) to (C6) to (C7);
  
  \node (C2) [blue!50!black] at (30+30*1.414,-30-30*1.414) {\small C$_2$};
  \draw (C7) to (C2);
  \node (C1) [blue!50!black] at (30,-30-60*1.414) {\small C$_1$};
  \draw (C1) to (C2);
  \node (C3) [blue!50!black] at (30+60*1.414,-30) {\small C$_3$};
  \draw (C3) to (C2);
  
  \node (H1) [blue!50!black] at (30-30*1.414,30+30*1.414) {\small H$_1$};
  \draw (C4) to (H1);
  \node (H2) [blue!50!black] at (30+90*1.414,-30+30*1.414) {\small H$_3$};
  \draw (C3) to (H2);
  \node (H3) [blue!50!black] at (30+90*1.414,-30-30*1.414) {\small H$_2$};
  \draw (C3) to (H3);
  \node (H4) [blue!50!black] at (30+60*1.414,-30-60*1.414) {\small H$_4$};
  \draw (C2) to (H4);
  \node (H5) [blue!50!black] at (30-30*1.414,-30-30*1.414) {\small H$_5$};
  \draw (C7) to (H5);

  \node (OH) at (30+30*1.414,-30-90*1.414) {\small OH};
  \draw (C1) to (OH);
  \node (O1) [red] at (30-30*1.414,-30-90*1.414) {\small O};
  \draw [double] (C1) to (O1);
  \node (O2) [red] at (-30-30*1.414,-30-30*1.414) {\small O};
  \draw [double] (C6) to (O2);
  \node (Cl1) [green!50!black] at (-30-30*1.414,30-30*1.414) {\small Cl};
  \draw (C5) to (Cl1);
  \node (Cl2) [green!50!black] at (-30-30*1.414,30+30*1.414) {\small Cl};
  \draw (C5) to (Cl2);
  \node (S) [purple] at (90,30) {\small S};
  \draw (C4) to (S) to (C3);

\end{tikzpicture}
\end{center}
\begin{center}
{\footnotesize\renewcommand{\arraystretch}{1.25}\setlength{\tabcolsep}{5pt}
\begin{tabular}{|c|D{.}{.}{1}D{.}{.}{1}D{.}{.}{1}D{.}{.}{1}D{.}{.}{1}D{.}{.}{1}D{.}{.}{1}D{.}{.}{1}D{.}{.}{1}D{.}{.}{1}D{.}{.}{1}D{.}{.}{1}|}
\hline
  $~~$ & \multicolumn{1}{c}{C$_1$} & \multicolumn{1}{c}{C$_2$} & \multicolumn{1}{c}{C$_3$} & \multicolumn{1}{c}{C$_4$} & \multicolumn{1}{c}{C$_5$} & \multicolumn{1}{c}{C$_6$}  & \multicolumn{1}{c}{C$_7$}  & \multicolumn{1}{c}{H$_1$} & \multicolumn{1}{c}{H$_2$} & \multicolumn{1}{c}{H$_3$} & \multicolumn{1}{c}{H$_4$} & \multicolumn{1}{c|}{H$_5$}\\
  \hline
 C$_1$ & 30020.09 &        &         &          &          &          &         &        &   &  &   &    \\
  C$_2$ & 57.58   & 8780.39  &         &          &          &          &         &        &   &  &     &  \\
  C$_3$ & -2.00   & 32.67    & 6245.45  &          &          &          &         &        &   &  &     &   \\
  C$_4$ & 0.02    & 0.30     & 0.00     & 10333.53  &          &          &         &        &   &  &     &     \\
  C$_5$ & 1.43   & 2.62     & -1.10     & 33.16     & 15745.40  &          &         &        &   &  &      &    \\
  C$_6$ &  5.54   & -1.66    & 0.00     & -3.53     & 33.16     & 34381.71     &      &        &   &  & &          \\
  C$_7$ &  -1.43   & 37.43    & 0.94     & 29.02     & 21.75     & 34.57     & 11928.71 &        &   &   &  &        \\
  H$_1$ &  0.04   & 1.47     & 2.03     & 166.60      & 4.06      & 5.39      & 8.61   & 3307.85 &   &   &  &        \\
  H$_2$ &  4.41   & 1.47     & 146.60   & 2.37      & 0.00      & 0.00      & 0.00     & 0.00    & 2464.15   &   &  &        \\
  H$_3$ &  1.86   & 2.44     & 146.60   & 0.04      & 0.00      & 0.00      & 0.00     & 0.18    & -12.41    & 2155.59  &  &        \\
  H$_4$ &  -10.10 & 133.60   & -6.97    & 6.23      & 0.00      & 5.39      & 3.80     & -0.68   & 1.28     & 6.00     & 2687.69   &        \\
  H$_5$ &  7.10   & -4.86    & 3.14     & 8.14    & 2.36      & 8.52      & 148.50    & 8.46    & -1.00     & -0.36    & 1.30    &  3645.08      \\
\hline
\end{tabular}
}
\end{center}
\caption{(a) Molecule structure of per-$^{13}$C-labeled dichlorocyclobutanone. (b) The system is divided into two subsystems. (c) The system is divided into four subsystems. (d) Hamiltonian parameters for the 12-spin NMR molecule.}
\label{molecule}
\end{figure}

\section{Test Example}
 
Our  demonstration example is a 12-spin molecule per-$^{13}$C labeled (1S,4S,5S)-7,7-dichloro-6-oxo-2-thiabicyclo[3.2.0]heptane-4-carboxylic acid. The molecular structure is shown in Fig. \ref{molecule}(a). The    molecular parameters including $\omega_i$ (diagonal) and $J_{ij}$ (off-diagonal) are given in Fig. \ref{molecule}(d).

In experiment, the experimental reference frequencies of the $^{13}$C channel and $^1$H channel are set to be $O_1 = 20696$ Hz and $O_2 = 2696$ Hz respectively. Let $\sigma_x$, $\sigma_y$, $\sigma_z$ denote the three Pauli operators. In the rotating frame, the system Hamiltonian takes the form 
\begin{equation}
H_S =    \sum_{i=1}^{12} {  \Omega_i   \sigma_z^i/2} + \pi\sum_{i<j}^{12} {J_{ij} \sigma_z^i \otimes \sigma_z^j/2},
\end{equation}
where $\Omega_i$ is the precession frequency of the spin $i$, $\Omega_i = -(\omega_i- O_1)$ for $i \le 7$ and $\Omega_i = -(\omega_i- O_2)$ for $i \ge 8$.  The control realized via applying an external radio-frequency field $u(t)=(u_x(t),u_y(t))$ ($0\le t \le T$), resulting in the control Hamiltonian
\begin{equation}
	H_C(t) = \sum_{i=1}^n \left( u_x(t) \sigma_x^i + u_y(t) \sigma_y^i \right).
\end{equation}
The controlled spin system's time evolution is then governed by the Schr{\"o}dinger equation
\begin{equation}
	\dot U(t) = -i [H_S + H_C(t)] U(t).
\end{equation}
The optimal control task is then, given a target operator $\overline U$, find a control pulse $u(t)$ such that the total time evolution propagator is as close to the target as possible.

%\subsection{Method I}
%

Previously, the idea of subsystem-based QOC has been used in Refs. \cite{Dawei17,PhysRevLett.123.030502} for obtaining high-fidelity pulses. In these references, the 12-spin system is divided into two subsystems: $A = \left\{ \text{C}_1,\text{C}_2,\text{C}_3,\text{H}_2,\text{H}_3,\text{H}_4\right\}$ and $B = \left\{ \text{C}_4,\text{C}_5,\text{C}_6,\text{C}_7,\text{H}_1,\text{H}_5\right\}$, each consisting of 6 spins; see Fig. \ref{molecule}(b). The only large couplings between these two subsystems are 
%$\left\{J_{\text{C}_1 \text{C}_6}, J_{\text{C}_1 \text{H}_5}, J_{\text{C}_2 \text{C}_5},J_{\text{C}_2 \text{C}_7},J_{\text{C}_2 \text{H}_5},J_{\text{C}_3 \text{H}_1},J_{\text{C}_3 \text{H}_5}, J_{\text{C}_4 \text{H}_2}, J_{\text{C}_4 \text{H}_4}\right\}$ 
$\left\{J_{\text{C}_1 \text{C}_6}, J_{\text{C}_1 \text{H}_5}, J_{\text{C}_2 \text{C}_7}, J_{\text{C}_4 \text{H}_4}\right\}$, so they can be approximately viewed as isolated. 
The 12-qubit optimal control problem can be treated as two 6-qubit problems.

\begin{figure*}
	\includegraphics{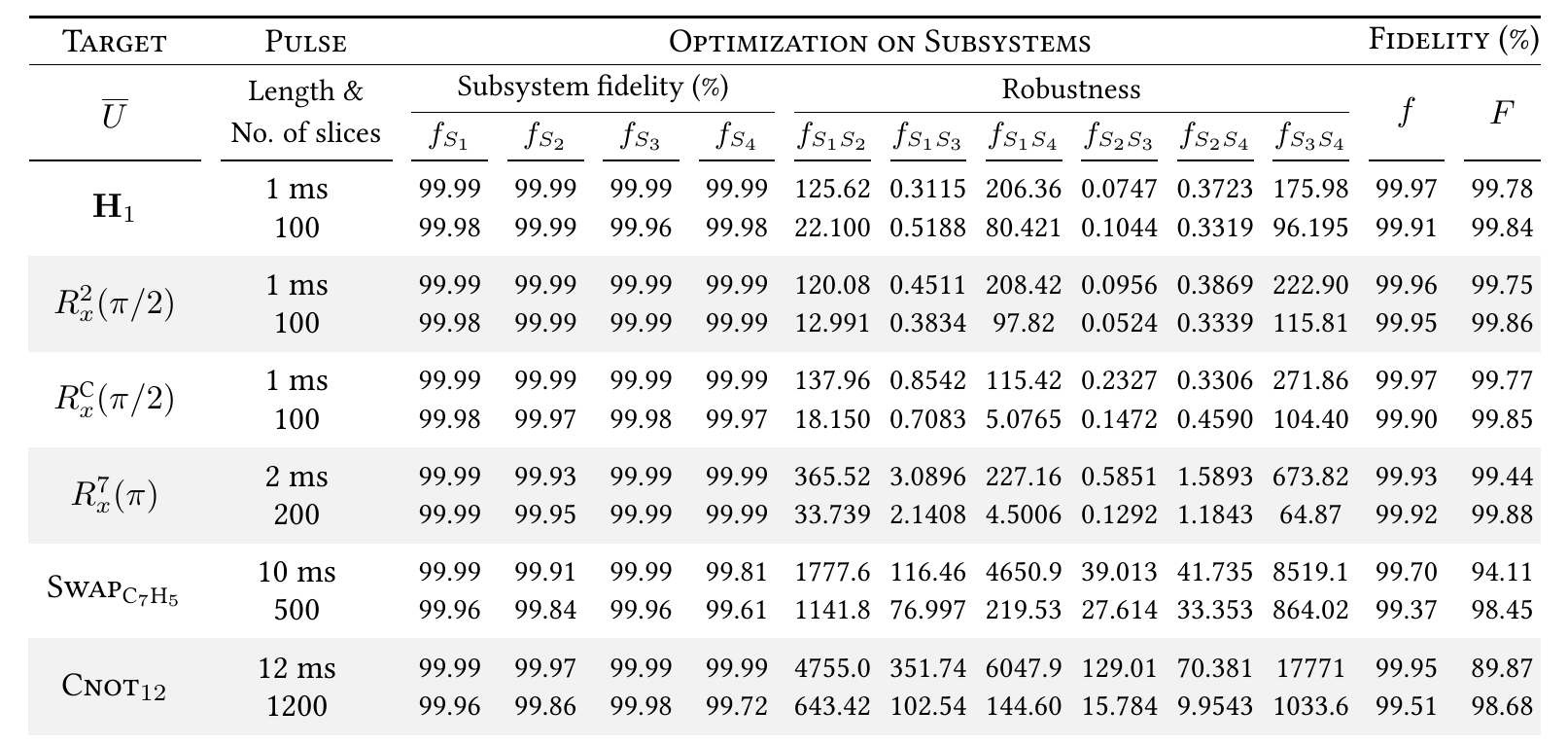}
	\label{results}
	\caption{Shaped pulse optimized by our subsystem-based QOC method. For each target operation, we start from a random pulse guess, first optimize its subsystem fidelities with only considering intra-subsystem Hamiltonian $H_0$, and then optimize its robustness to inter-subsystem Hamiltonian $H_1$. In each row, the upper data mean pulse fidelities and robustness before the stage of robustness optimization, and the lower data mean pulse fidelities and robustness after the stage of robustness optimization.}
\end{figure*}

However, as we have mentioned in our main text that, the above strategy actually lacks a solid theoretical grounding.   As we have derived from the general framework, to solve the robustness problem  (to first order), we have to consider pairs of subsystems, rather than on solely individual subsystems. Only in this way could it be guaranteed that pulses with high subsystem fidelities also has high fidelity on the full system. We have tested our improved subsystem-based QOC algorithm on the 12-qubit system. The results are shown in Fig. \ref{results}.

\end{document}